\documentclass[twocolumn,prl,superscriptaddress]{revtex4}
\usepackage{graphicx}
\usepackage{dcolumn}% Align table columns on decimal point
\usepackage{bm}% bold math
\usepackage{amsmath}
\usepackage{times}
\usepackage{epstopdf}
\usepackage{color}
\usepackage[breaklinks=true,colorlinks,citecolor=blue,linkcolor=blue,urlcolor=blue]{hyperref}

\makeatletter

\newcommand{\Rmnum}[1]{\expandafter\@slowromancap\romannumeral #1@}
\makeatother
\graphicspath{{images/}}
\begin{document}

\title{Charge density wave and Weyl Semimetal phase in Y$_2$Ir$_2$O$_7$}
%\title{Sliding density wave in low dimensional single crystalline Pyrochlore Iridates}
\author{Abhishek Juyal}
%\email{abijuyal@iitk.ac.in}
\thanks{These authors contributed equally}
\affiliation{Department of Physics, Indian Institute of Technology Kanpur, Kanpur 208016, India}
\affiliation{Georgia Tech Lorraine, IRL 2958-CNRS, 57070 Metz, France}
\author{Vinod Kumar Dwivedi}
\thanks{These authors contributed equally}
\affiliation{Materials Science Program, Indian Institute of Technology Kanpur, Kanpur 208016, India}
\affiliation{Department of Physics, Indian Institute of Science, Bengaluru 560012, India}
\author{Sonu Verma} 
\affiliation{Department of Physics, Indian Institute of Technology Kanpur, Kanpur 208016, India}
\author{Shibabrata Nandi}
\affiliation{Department of Physics, Indian Institute of Technology Kanpur, Kanpur 208016, India}
\affiliation{Forschungszentrum J\"ulich GmbH, J\"ulich Centre for Neutron Science (JCNS-2) and Peter Gr\"unberg Institut (PGI-4), JARA-FIT, 52425 J\"ulich, Germany }
\affiliation{RWTH Aachen, Lehrstuhl f\"ur Experimentalphysik IVc, J\"ulich-Aachen Research Alliance (JARA-FIT), 52074 Aachen, Germany}
\author{Amit Agarwal}
\email{amitag@iitk.ac.in}
\affiliation{Department of Physics, Indian Institute of Technology Kanpur, Kanpur 208016, India}
\author{Soumik Mukhopadhyay}
\email{soumikm@iitk.ac.in}
\affiliation{Department of Physics, Indian Institute of Technology Kanpur, Kanpur 208016, India}

\begin{abstract}
The subtle interplay of band topology and symmetry broken phase, induced by electron correlations, has immense contemporary relevance and potentially offers novel physical insights. Here, we demonstrate charge density wave (CDW) in bulk Y$_2$Ir$_2$O$_7$ for $T<10$ K, and its transition to the Weyl semimetal (WSM) phase at higher temperatures. The CDW phase is evidenced by a) current induced nonlinear conductivity with negative differential resistance at low temperature, b) low frequency Debye like dielectric relaxation at low temperature with a large dielectric constant $\sim 10^8$, and c) an anomaly in the temperature dependence of the thermal expansion coefficient. The WSM phase at higher temperature is confirmed by the DC and AC transport measurements which show an inductive response at low frequencies. More interestingly, we show that by reducing the crystallite size, the low temperature CDW phase can be eliminated leading to the restoration of the WSM phase. 
\end{abstract}

%\pacs{75.47.Lx, 72.20.-i, 79.60.-i, 71.70.Ej, 78.70.Dm}

\maketitle

%{\it Introduction.---} 
Strong electron-electron correlation and topology are generally considered mutually exclusive domains of physics. However, recent observations of interaction driven topological phase transition~\cite{Shi, Polshyn} have opened up prospects for exploring the interplay of correlation and topology. In fact, Weyl semimetals (WSM) were first predicted in correlated pyrochlore iridates, R$_2$Ir$_2$O$_7$ (R = Y, Eu, Nd)~\cite{Wan}. These $5d$ orbital based iridates can host a variety of topological and quantum phases in addition to the WSM phase \cite{Wan,Kim,Pesin,Kargarian,Machida,Yang,William,Kalu,Dwivedi, Bikash1, Bikash2}. The search for Fermi arc surface states in Y$_2$Ir$_2$O$_7$ and other iridates has been unsuccessful due to the unavailability of good quality single crystals and inadequate surface preparation rendering techniques such as ARPES and STM, ineffective. However, there could be a more fundamental reason: Weyl nodes could gap out forming density wave instabilities due to chiral symmetry breaking induced by coulomb interaction~\cite{Axion-CDW, Abhishek1, Abhishek2}. Recently, a large negative contribution to the longitudinal magnetoresistance in the sliding mode of the charge density wave (CDW) phase of WSM candidates including nanorods based on Y$_2$Ir$_2$O$_7$ have been observed~\cite{Abhishek2, Gooth}. The negative longitudinal magnetoresistance in such cases originates from the axionic contribution of the chiral anomaly to the phason current~\cite{Abhishek2, Gooth}. 
%In the absence of the effectiveness of direct experimental probes, it is imperative to try out alternate experimental routes for exploring the WSM phase in bulk Y$_2$Ir$_2$O$_7$. 

\begin{figure}
\includegraphics[width=0.99\linewidth]{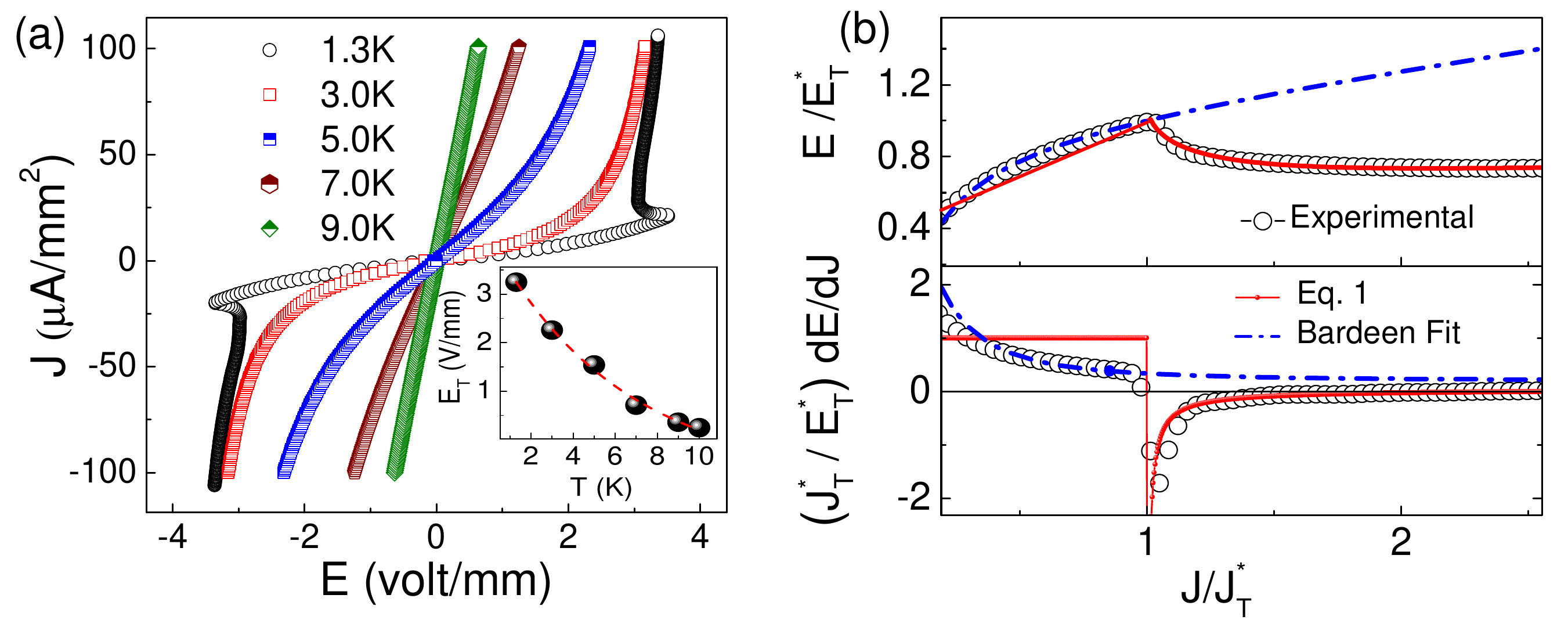}
\caption{CDW phase in Y$_2$Ir$_2$O$_7$ for $T < 10$ K. a) DC IV characteristics for bulk Y$_2$Ir$_2$O$_7$ showing CDW de-pinning induced nonlinear transport below 9 K, and S-shaped negative differential resistance at 1.3 K. 
The inset shows de-pinning threshold ($E_{\rm T}$) for positive bias, which follows $E_{\rm T} \propto \exp[-T/T_0]$, with $T_0 \approx 3$ K.
b) Characteristic plot (at $T=1.3$ K) of $E$ (top) and $dE/dJ$ (bottom) normalized to $E_{\rm T}$ and $J_{\rm T}$ where current is regulated. The continuous lines are the theoretical fits to Bardeen's tunneling theory in pinned CDW for $ J \in [J_{\rm T}, J_{\rm T}^*]$ 
and to Eq.~\eqref{Eq11} for $J > J_{\rm T}^*$ in the de-pinned regime. 
\label{IV}}
\end{figure}

Apart from Y$_2$Ir$_2$O$_7$, amongst the iridates family, Nd$_2$Ir$_2$O$_7$ shows a quadratic band touching at the $\Gamma$ point which gaps out at low temperature~\cite{Nakayama}. The claim of metal-semimetal transition in the optical conductivity data for Eu$_2$Ir$_2$O$_7$~\cite{Sushkov} needs to be backed by other more compelling evidence. Epitaxial strain induced `all-in, all-out' ordering in thin films of Pr$_2$Ir$_2$O$_7$ breaks the time reversal symmetry, possibly leading to WSM phase~\cite{Yangyang}. Although single crystalline nanorods of Y$_2$Ir$_2$O$_7$ have shown evidence of possible chiral anomaly in the gapped out WSM  phase~\cite{Abhishek1, Abhishek2}, the question remains whether the conclusions drawn can be extended to bulk Y$_2$Ir$_2$O$_7$ as well. 

Here, we present experimental evidence of gapped out WSM ground state in bulk Y$_2$Ir$_2$O$_7$, showing characteristics of (axionic) CDW with a distinct transition to WSM phase above $10$ K. Both the CDW phase in bulk Y$_2$Ir$_2$O$_7$ for $T<10$ K, and the WSM phase above $T>10$ K are confirmed independently via DC and AC transport measurements. The WSM to CDW transition is also observed in the thermal expansion experiments. The CDW gap opening can be prevented by reducing the grain size, thus extending the WSM state to low ($\sim 1$ K) temperatures. Remarkably, we find the DC transport in nano-crystalline samples to be governed by the Coulomb interaction induced diffusive transport in the WSM phase \cite{Pavan}.

%%{\it CDW phase in Bulk Y$_2$Ir$_2$O$_7$.---} 
Electrical transport measurements were performed on bulk and nano-crystalline Y$_2$Ir$_2$O$_7$ samples. See supplementary material (SM) \cite{SM} for details of sample preparation and characterization.  % and in Ref.~\cite{Dwivedi}. 
To start with, we present the evidence for the CDW phase in the bulk sample, that is observed in {\it current-driven} nonlinear IV characteristics shown in Fig.~\ref{IV}. The linear conductivity at low voltage is followed by the onset of nonlinear conduction at higher voltages when the CDW starts sliding. More interestingly, the non-ohmic IV characteristics at 1.3 K also exhibits current controlled negative differential resistance (NDR). Both of these effects are associated with CDW de-pinning, and the resultant constant current IV characteristic showing NDR is commonly referred to as `\textit{S}-shaped' response. The NDR is not observed at higher temperatures although the nonlinear IV characteristic is found to persist up to $9$ K in  Fig.~\ref{IV}(a). 

\begin{figure}
\includegraphics[width=0.8\linewidth]{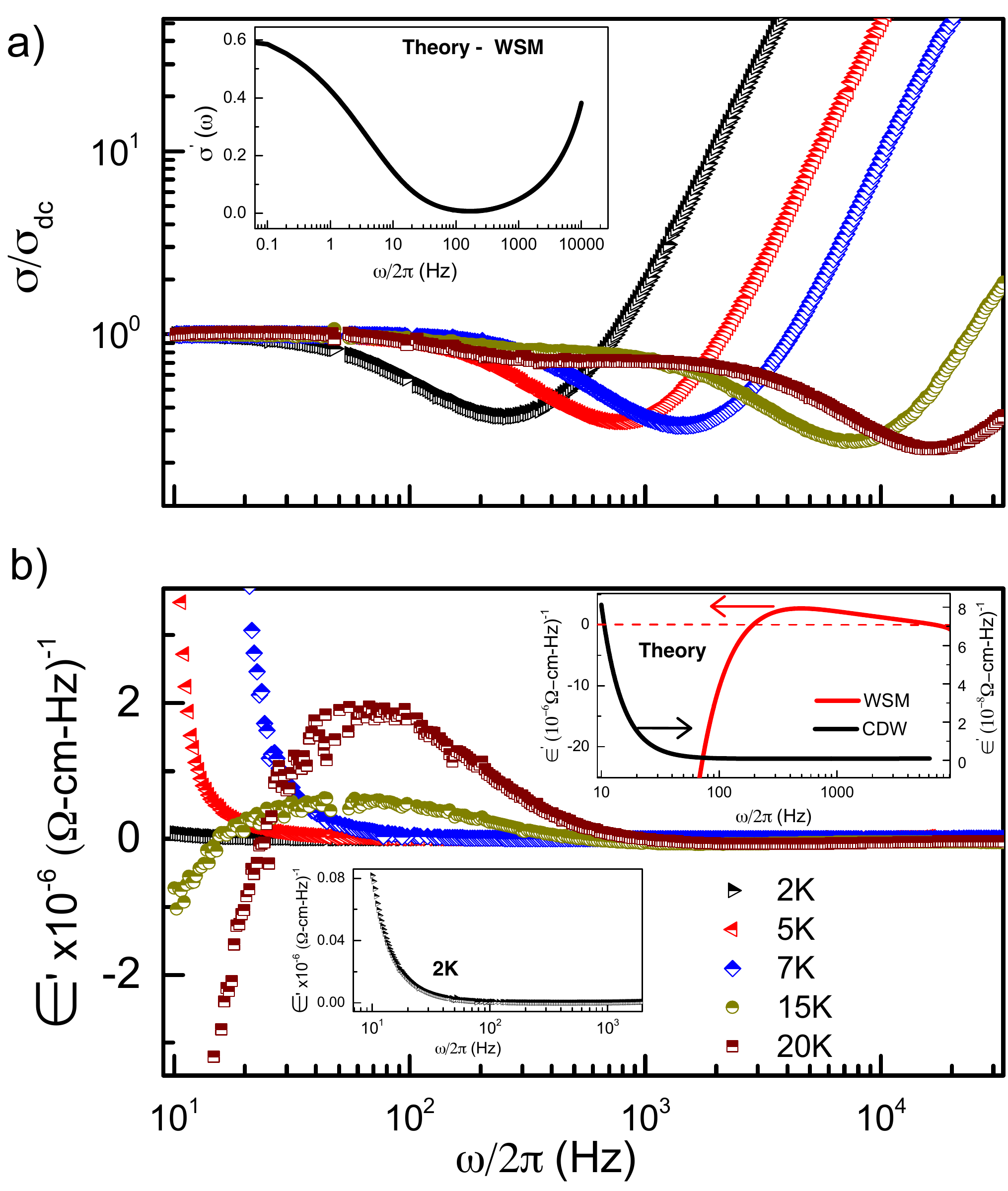}
\caption{a) $\rm{Re} [\sigma(\omega)]$ %frequency dependence of the real part of conductivity 
for the {\it bulk polycrystalline samples} shows a WSM like 
behaviour for $T > 10$ K, while for $T < 10$ K it shows a CDW behaviour with Debye like relaxation. The inset shows the theoretically calculated conductivity for disordered WSM. %% which resembles the measured conductivity for $T>10$ K. b) %The contrasting behaviour of the CDW state and the WSM phase is more marked in the plot of the real part of the dielectric function. 
(b) The real part of dielectric function for $T<10$ K is consistent with Debye relaxation (see black curve in the top inset). The bottom inset shows the blown-up portion of the experimental plot at $2$ K. In contrast, the WSM state for $T>10$ K shows a crossover from negative to positive dielectric constant, consistent with the theoretical results for  WSM, shown by the red curve in the top inset.
\label{lowfreq}}
\end{figure}

\begin{figure*}[t]
\includegraphics[width=0.8\linewidth]{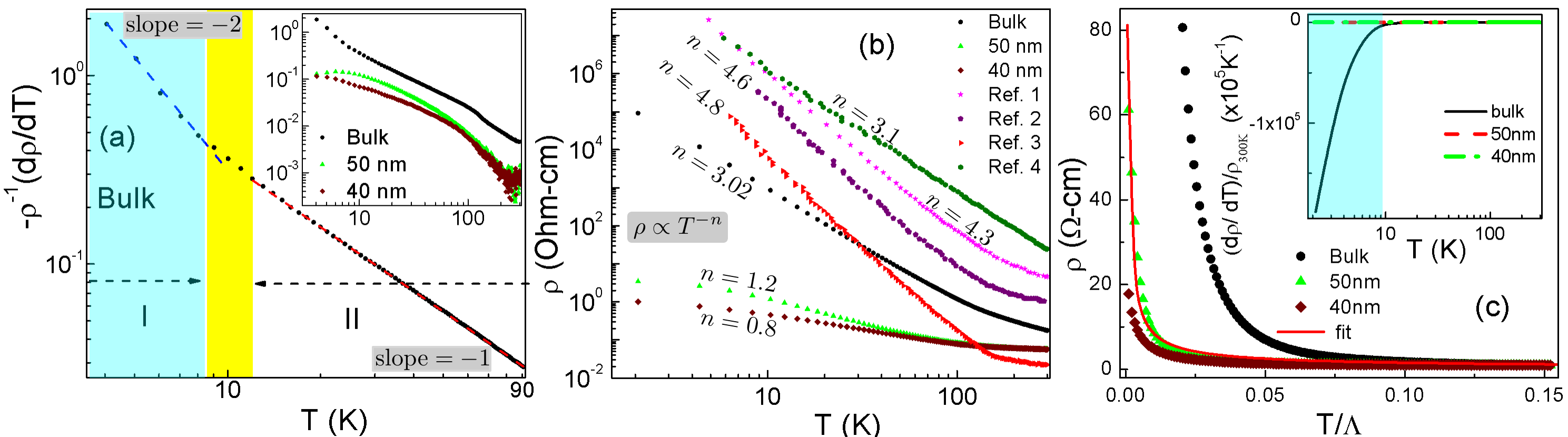}
\caption{ $\rho(T)$ of bulk poly-crystalline sample. (a) $\rho^{-1} \frac{d\rho}{dT}$ plotted as a function of $T$ clearly shows an activated behaviour (with slope $\approx -2$) for $T < 10$ K and a power law  (with slope $\approx -1$) above $T > 10$ K. The inset shows the corresponding $\rho -T$ behaviour of the bulk and nano-crystalline samples. (b) Comparison of the power law exponents from different experiments with that for our bulk, and nano-crystalline samples. (c) The variation of the DC conductivity of bulk and the nano-crystalline (average particle size of 40nm and 50nm) samples with  temperature, compared with the Coulomb interaction induced diffusive transport model (continuous line) for interacting WSM. 
%with $24$ isotropic Weyl nodes. 
Inset: Plot of $\frac{d\rho}{dT}$ vs $T$ clearly shows the resistive anomaly around $T_{\rm CDW}$ for the bulk sample.%, in contrast to the nano-crystalline samples. 
%Here, $\Lambda$ is the energy cutoff for the linear bands. 
\label{fig4}}
\end{figure*}

The essential features of these measurements are captured by a simple model describing the dynamics of the CDW phase. The CDW phase may be weakly pinned by impurities, lattice defects or confined by the grain boundaries~\cite{Dubiel}. Electrical transport occurs due to the translational motion of the CDW and that of the normal carriers. At low electric field (or current) values below a critical threshold, $E<E_{\rm T}$ (or $J < J_{\rm T}$), normal quasiparticles carry the current leading to ohmic behaviour while the CDW domains remain pinned. For $E > E_{\rm T}$ (or $J > J_{\rm T}$), the CDW domains become partially de-pinned, leading to nonlinear contribution to conductivity. In this regime, the transport is  described by Bardeen's CDW tunneling theory \cite{Bardeen}. The threshold field directly probes the effectiveness of the pinning of the CDW order parameter. We find $E_{\rm T}(T) = E_0 \exp(-T/T_0)$, with $T_0 \approx 3$ K denoting the strength of the pinning potential \cite{Maki}, as shown in the 
inset of Fig.~\ref{IV}(a). 

On increasing the current (or field) beyond a second threshold, $J > J_{\rm T}^*$ (or $E > E_{T}^*$), the CDW dislodges and moves with a single phase. Consequently, for  $J > J_{\rm T}^*$ the voltage across the sample drops abruptly as the current increases. 
This leads to NDR as shown in Fig.~\ref{IV}(a) for $T=1.3$ K. In this regime, the conductivity is dominated by coherent CDW transport, which is well described by the damped dynamics of the CDW phase \cite{CDW_RMP,GG}. 
In a constant current experiment, the nonlinear I-V characteristics {\it above} the de-pinning threshold $E_{\rm T}^*$ is given by \cite{Monceau,CDW_RMP}
\begin{equation} \label{Eq11}
 E = \frac{J}{\sigma_n}-\frac{\beta E_{\rm T}^*}{1+\beta} \left[ \left( \frac{J}{J_{\rm T}^*} \right) ^2 - 1 \right]^{1/2}~.
\end{equation}
Here $\sigma_n$ is the ohmic conductivity of the normal quasiparticles, and $\beta$ is a system and temperature dependent parameter. The conductivity $dE/dJ$ has a discontinuity at the NDR threshold $J_T^*$. This discontinuity is explicitly shown for the $T= 1.3$ K curve in both panels of Fig.~\ref{IV} (b), along with the corresponding fit to Bardeen's tunneling theory for $J_{\rm T} < J < J_{T}^*$ and to Eq.~\eqref{Eq11} for $J > J_{\rm T}^*$. 
This is possibly the first demonstration of the coexistence of quantum and classical regime of the CDW transport. 
{Interestingly, the IV characteristics in the NDR regime also show negative (positive) longitudinal (transverse) magneto-resistance, similar to that found in single crystalline nanorods of Y$_2$Ir$_2$O$_7$ \cite{Abhishek2}. This is shown explicitly in Fig.~S3 in SM \cite{SM}.

We measure the low-frequency dielectric response using the standard lock-in technique. 
For $T< 10$ K, the AC transport measurements show distinctly different behaviour compared to the measurements at higher temperatures, as highlighted in Fig.~\ref{lowfreq}. In particular, we find that for $T<10$ K, the dielectric constant shows a Debye like relaxation indicating a CDW phase. 
Similar to the observations in single crystalline Y$_2$Ir$_2$O$_7$ nanorods \cite{Abhishek1,Abhishek2}, 
we find that the DC conductivity and the peak frequency ($\omega_p$) of the imaginary part of the dielectric constant show an Arrhenius-like temperature dependence: 
$\sigma_0 \propto \omega_p \propto \exp[-\Delta/T]$, with $\Delta = 16.1$ K. Physically this occurs due to the screening of the damped collective charge oscillations of the CDW by the thermally excited normal carriers \cite{Blumberg,Tucker1}. This gives us a mean-field estimation of the 
CDW transition temperature ( $2\Delta=3.52~T_{\rm CDW}$)  to be $T_{\rm CDW}\approx9.15$~K, consistent with the DC measurements. 
The corresponding coherence length can be estimated to be $\xi_0\approx 2.4$ nm, which is much smaller than the grain size$\sim 1~\mu$m, and similar to the observed value in other poly-crystalline samples~\cite{Yaoa}.

For $T> 10$ K, the real component of the conductivity [see Fig.~\ref{lowfreq}(a)] displays behaviour similar to that of the Weyl semimetal phase with short range disorder \cite{Pavan} [see inset in Fig.~\ref{lowfreq}(a)]. 
The conductivity remains almost constant for very low frequencies ($\omega < 10^2$ Hz), decreases for intermediate frequencies ($\omega \in[10^2,10^3]$ Hz), and then increases very rapidly for $\omega \approx 5 \times 10^4$ Hz, depending on the temperature. 
Within the Born approximation, the conductivity in WSM with dilute and random short range disorder is given by \cite{Pavan}
\begin{equation} \label{JJ}
\sigma(\omega, T) = \frac{e^2 v_F^2}{h \gamma} J\left(\frac{N\omega}{\omega_0}, \frac{N T}{\omega_0}\right). 
\end{equation}
Here, $\omega_0 = 2 \pi v_F^3 /\gamma$ is the characteristic 
frequency set by the disorder scale $\gamma$, and $N$ denotes the number of Weyl nodes. In Eq.~\eqref{JJ}, we have defined 
$J\left(\bar \omega, \bar T\right) = \frac{4}{3}\int\frac{d {\bar \epsilon}}{2\pi} \frac{[f_{\bar{T}}(\bar{\epsilon})-f_{\bar{T}}(\bar{\epsilon}+{\bar \omega})]}{\omega}  I(\bar{\epsilon}+{\bar \omega,\bar{\epsilon} })$~,
%\end{equation}
with $\bar x = x/\omega_0$, $f_{\bar{T}}(\bar \omega) = (1+e^{\bar \omega/\bar T})^{-1}$ is the Fermi function. The explicit form of the function $I(\bar{\epsilon}+{\bar \omega},\bar{\epsilon})$ is given in the SM \cite{SM}. 
Equation~\eqref{JJ} implies that $\sigma(\omega)$ diverges as $\omega \to \omega_0$, signalling a breakdown of the Born approximation \cite{Pavan}. Thus, we estimate that $\omega_0\approx 4 \pi \times 10^4$ rad/s, for the $T =20$ K curve in Fig.~\ref{lowfreq}. 
In the $\omega \ll T$ limit (valid for our experiments), Eq.~\eqref{JJ} leads to $\sigma (\omega \to 0) \approx 2 e^2 \omega_0/(3 \hbar v_F)$. Using the 
estimated value of $\omega_0$, we calculate $\sigma (\omega \to 0) = 5.2\times 10^{-5}/(\Omega$ cm), which is reasonably close to the %5.2\times 10^{-6}/(\Omega$ cm), which is reasonably close to the
measured value of $\sigma (\omega \to 0) = 3.2\times 10^{-5}/(\Omega$ cm).
This non-monotonic behaviour in the low frequency conductivity for $T>10$ K is also accompanied by a non-monotonic behaviour in the dielectric constant, as shown in Fig.~\ref{lowfreq}(b). The real part of the dielectric constant is negative for low frequencies, it increases to become positive at intermediate frequencies, and then decreases again to 
settle just below the zero axis, consistent with the theoretical calculation shown in the inset of Fig.~\ref{lowfreq}(b). 
In the CDW regime for $T < 10$ K, the dielectric function shows a monotonically decreasing behaviour with frequency, similar to that seen in other CDW systems, including crystalline nanorods of Y$_2$Ir$_2$O$_7$ \cite{Abhishek1,Abhishek2}. Remarkably, the CDW transition temperature in Y$_2$Ir$_2$O$_7$ drops from $T_{\rm CDW} > 300$ K in single crystalline nanorods \cite{Abhishek2} to $T_{\rm CDW} \approx 10$ K in bulk polycrystalline samples. 

The CDW to WSM transition in Y$_2$Ir$_2$O$_7$ can also be seen from the temperature dependence of the DC conductivity. To distinguish the nature of DC transport between 
activated and power law behaviour,  
we plot $-\rho^{-1} \frac{d\rho}{dT}$ as a function of $T$ on a log-log scale for bulk sample in Fig.~\ref{fig4}(a). 
The activated behaviour of the DC transport for $T < 10$ K (with a slope $\approx -2$ in the CDW phase) and a power law behaviour above $T > 10$ K (with a slope $\approx -1$ in the WSM phase) is clearly established. The CDW transition temperature estimated by fitting the activated DC transport 
[$\rho \propto e^{\Delta_{\rm CDW}/(k_B T)}$] below $T < 10$ K is consistent with the earlier estimated value. 

For $T > 10$ K, a clear power law behaviour in the resistivity,  $\rho(T) \propto T^{-n}$, is found till $T \approx 80$ K. However, different experiments have reported 
varying power-laws for the resistivity of Y$_2$Ir$_2$O$_7$. While our bulk samples show $n=3.02$, close to the value of $n =3.1 $ reported in \cite{Kumar2017}, there have also been reports of $n = 4.3$  \cite{Ramirez}, $n = 4.6$ \cite{Dissler2012} and $n = 4.8$ \cite{Liu2014}. See Fig.~\ref{fig4}(b) for a comparison, where we estimate the exponents by digitizing and fitting all the available data-sets up to $50$ K, for uniformity. This clearly shows that the low temperature power-law behaviour of $\rho(T)$ in Y$_2$Ir$_2$O$_7$ is {\it non-universal}, and possibly dependent on various factors such as crystallite size, disorder layout, etc. 

The nano-crystalline samples show a power law exponent close to unity, which has also been predicted to arise in the Coulomb interaction dominated diffusive transport regime of WSM. % near the charge neutrality point. 
The predicted DC conductivity \cite{Pavan} is given by 
\begin{equation}
\sigma_{dc}(T)=\frac{e^2}{h}\frac{k_B T}{\hbar v_F(T)}\frac{0.45}{\alpha_T^2ln\alpha_T^{-1}}~.
\end{equation}
Here, the renormalized fine structure constant is $\alpha_T=\alpha_1[1+\frac{(N+2)\alpha_1}{3\pi}ln\frac{(\hbar\Lambda)}{k_B T}]$, and the renormalized Fermi velocity is $v_F(T)=v_F(\alpha_1/\alpha_T)^{2 + 2/N}$. The number of Weyl nodes is specified by $N=24$, and $\Lambda$ is a momentum cutoff set by the separation
between the Weyl nodes.
%For the clean WSM state in Y$_2$Ir$_2$O$_7$, we have the $v_F=2\times10^5$ m/s, $\alpha_1=0.5$, $N=24$ and $\Lambda=0.17$eV~\cite{Pavan}. 
This model description for $\rho(T)$ \cite{Pavan} fits our experimental data for the nano-crystalline sample, remarkably well {\it over the entire temperature range}, as shown in Fig.~\ref{fig4}(c). 

This naturally prompts the question: why does the CDW gap close on reducing the particle size (or increasing the surface to volume ratio), leading to the restoration of the WSM phase in Y$_2$Ir$_2$O$_7$~\cite{Catherine}? The possibility of the quantum confinement gap [$\Delta_{\rm QC} \approx {\hbar^2/(m_e L^2)}$] becoming larger than the CDW gap is unlikely in this case as, even for an average particle size of $50$nm we have $\Delta_{\rm QC} \sim 0.03$ meV, which is almost 30 times smaller than $k_B T_{\rm CDW} \approx 0.86$ meV. We rule out the possibility of the softening of phonon mode due to increased surface contribution by our specific heat measurements (not shown here), which shows an enhancement in the $\theta_D$ in the nano-crystalline sample ($\theta_D=503.9\pm14.5$ K) compared to the corresponding bulk value ($\theta_D=462.6\pm9.1$ K). 
One possibility supported by our X-ray spectroscopy data is the increase in carrier kinetic energy in nano-crystalline samples owing to the enhancement of the concentration of higher oxidation states - Ir$^{5+}$ (See Fig. S2 in SM) or due to quantum confinement effects. The increased kinetic energy in nano-crystalline samples is likely to reduce the impact of correlation effects and reduce the $T_{\rm CDW}$. Recently pressure induced closing of a large CDW gap has been observed in Ta$_2$Se$_8$I, which exhibits a Weyl semimetal to CDW transition close to room temperature in ambient pressure~\cite{Mu}. In our case, the CDW gap being much smaller, increased surface pressure due to reduction of particle size could be effectively playing a similar role. 

\begin{figure}
\includegraphics[width=0.7\linewidth]{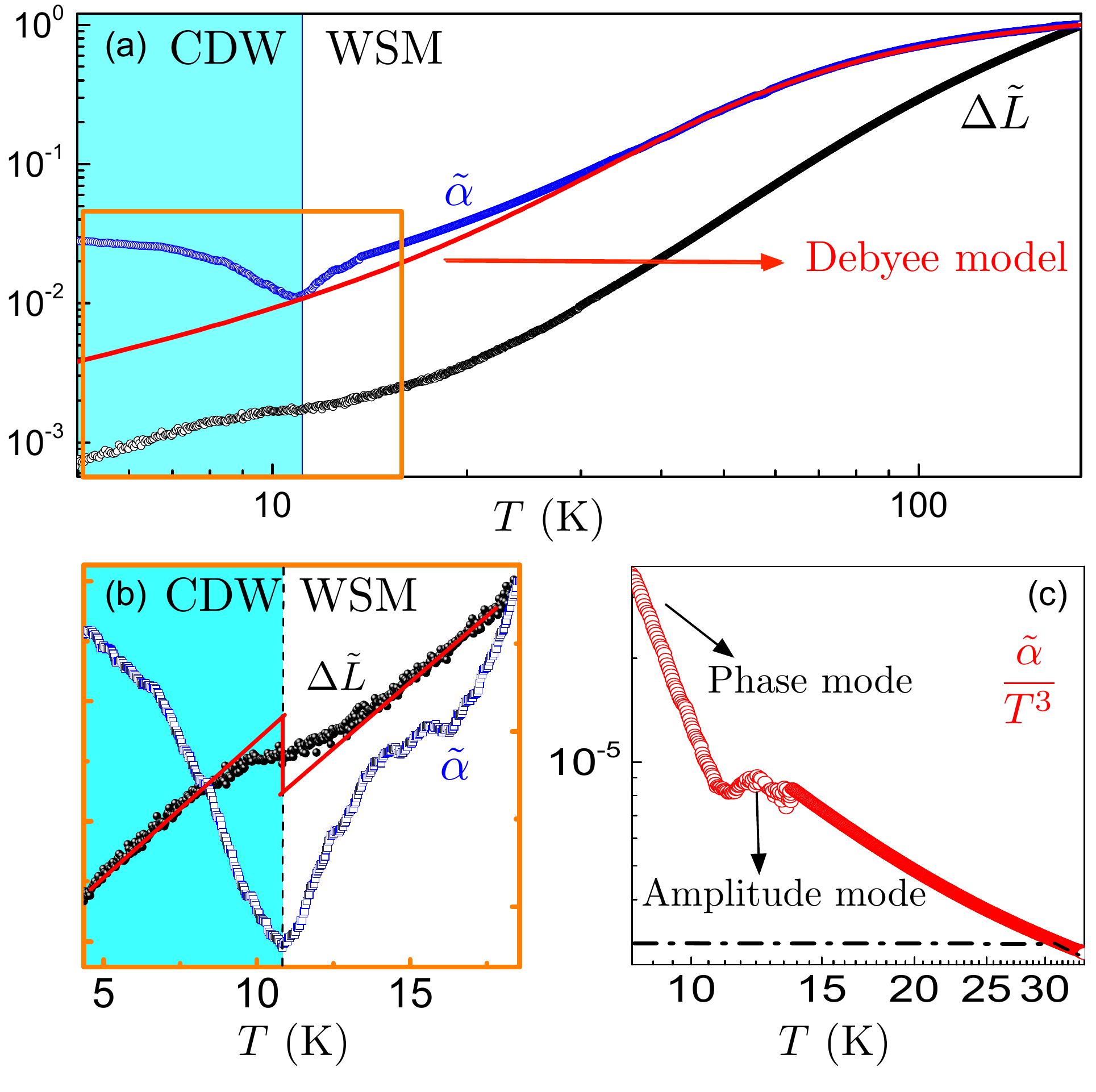}
\caption{a) Variation of the relative length ($\Delta {\tilde L} = \Delta L/\Delta L_{T=180}$) and $\alpha$ (${\tilde \alpha} = \alpha/\alpha_{T=180}$) with $T$ for the bulk polycrystalline Y$_2$Ir$_2$O$_7$. % in absence of magnetic field.
The thermal expansion coefficient (blue curve) fits well with the Debye model (red curve) at high $T$. However, it shows an anomaly at low temperature coinciding with the $T_{\rm CDW}$ obtained from the transport measurements. This is highlighted in the zoomed panel b) showing the low temperature regime. Here, we have shifted the $\Delta {\tilde L}$ curve and the different linear fits below and above $T_{\rm CDW}$ are shown by red lines. 
c) The deviation of $\alpha/T^3$ from the Debye behaviour (dashed black line) at low $T$. \label{fig3}}
\end{figure}

For independent confirmation of the CDW phase transition in Y$_2$Ir$_2$O$_7$, we performed thermal expansion measurements using a capacitive dilatometer having sub-angstrom resolution with high sensitivity of $\Delta L/L \sim 10^{-10}$ at low temperatures. We measured the linear thermal expansion coefficient, $\alpha (T) \equiv dL/dT$, and verified it for different heating and cooling rate.
%ranging between 0.3 K/min and 1.5 K/min. 
%The thermal expansion data on bulk Y$_2$Ir$_2$O$_7$ are presented in Fig.~\ref{fig3}. 
The relative thermal expansion ${\Delta L}/{L}$, shown in Fig.~\ref{fig3}, decreases monotonically with the lowering of temperature. %, as expected in cases \textcolor{red}{where the contribution is only due to the lattice}. 
However, there is an anomaly at $T \approx 10.4$ K (consistent with the estimated $T_{\rm CDW} \approx 9.15$ K from transport measurements) with relative dilatation of about $1.02\times10^{-6}$, as shown in Fig.~\ref{fig3} (a)-(b). The anomaly is highlighted further in the thermal expansion coefficient which shows a sharp minimum at the same temperature. The $\alpha(T)$ curve fits reasonably well to the Debye model at higher temperature (see SM \cite{SM} for details).
%, with 
%\begin{equation}{\label{Eq1}}
%\alpha(T) = \alpha_0\bigg( \frac{T}{\Theta_D}\bigg)^{3} \int_{0}^{\Theta_D/T} \frac{x^4 e^x}{(e^x-1)^2} dx~.
%\end{equation}
%Here, $\alpha_0$ is a temperature independent fitting parameter and $\Theta_D$ is the Debye temperature. Our 
%the estimated value of $\Theta_D =339$ K. This agrees well with earlier specific heat based measurements in Y$_2$Ir$_2$O$_7$~\cite{Taira}. 
The deviation from the Debye $T^3$ law in $\alpha (T)$ is highlighted by plotting the residual $\alpha(T)/T^3$ vs $T$ on the log-log scale in Fig.~\ref{fig3} (c). Similar deviation from the $T^3$ law has also been observed in low temperature specific heat for other CDW systems
\cite{Biljakovic,Cp_Anomaly2,Cp_Anomaly3}. This deviation can be attributed to the splitting of the acoustic phonon mode (for $T < T_P$) into the gapped amplitude mode [which gives the `hump' for $T \sim T_P$ in $\alpha(T)/T^3$ plot in Fig.~\ref{fig3}(c)] and the gapless phase mode ($T\to 0$ behaviour of the $\alpha(T)/T^3$ curve) of the CDW condensate.

To summarize, we present clear experimental evidence of temperature and grain-size dependent phase transition from the CDW phase to the WSM phase in bulk Y$_2$Ir$_2$O$_7$ using DC and AC transport experiments. The phase transition is independently confirmed via thermal expansion measurements as well. The low frequency dielectric response in the WSM phase at high temperature is consistent with the theoretically predicted response in WSM having short range neutral disorder. %with renormalized Fermi velocity. % strongly renormalized by the Coulomb interactions. 
More interestingly, we demonstrate that the CDW gap is suppressed significantly on reducing crystallite size.  %and it almost vanishes in nano-crystalline samples with grain size less than 50 nm. 
The observed DC conductivity over a wide temperature range in the nano-crystalline Y$_2$Ir$_2$O$_7$ shows interaction induced diffusive behaviour characteristic of WSM. %For bulk samples, we show that the power law expo
%Our transport experiments clearly establish the existence of the CDW and the WSM phase in bulk Y$_2$Ir$_2$O$_7$. 
The grain size dependent tunability of the phase transition clearly suggests the existence of a quantum critical point separating the WSM phase and the broken symmetry (possibly axionic) CDW phase. 
This opens up new directions for exploring quantum criticality and interaction driven interplay between topological and symmetry broken phases.

\section*{Acknowledgements}
We acknowledge Science Engineering and Research Board (SERB) and Department of Science and Technology (DST) of government of India for financial support.

%ground state with reduction in grain size in Y$_2$Ir$_2$O$_7$. Although we do not have, at present, a theory that incorporates both interaction and disorder, it seems that the relaxation rate in the low frequency dielectric response of the disordered Weyl semimetal is strongly re-normalized by interaction effect. The interplay of density wave and WSM phase in the present work combined with our earlier works on single crystalline Y$_2$Ir$_2$O$_7$ nanowires clearly suggests existence of two possible topological phases in correlated Iridates: a WSM phase or a gapped out WSM phase in the form of axionic CDW.

\end{document}